\documentstyle[12pt]{article}

\newcommand{\bq}{\begin{equation}}
\newcommand{\eq}{\end{equation}}

\begin{document}
\sloppy
\thispagestyle{empty}
%\begin{flushleft}
%DESY 95--175 \\
%October 1995\\
%\end{flushleft}

\setcounter{page}{0}

\mbox{}
\vspace*{\fill}
\begin{center}
{\large\bf  REVIEW OF HIGHER ORDER QCD CORRECTIONS
                         TO STRUCTURE FUNCTIONS}
\\
\vspace{4em}
\large
                         W.L. van Neerven
\\
\vspace{4em}
\normalsize

{\it   Instituut-Lorentz, University of Leiden, P.O. Box 9506,            
                           2300 RA Leiden, The Netherlands}\\
\end{center}
\vspace*{\fill}
\begin{abstract}
\noindent
  A review is presented on all higher order QCD corrections to deep
  inelastic structure functions. The implications of these corrections
  for polarized and unpolarized deep inelastic lepton-hadron scattering
  will be discussed.\\[6cm]
To appear in the proceedings of the 1996 HERA Physics Workshop
\end{abstract}
\vspace*{\fill}
\newpage
%
%%%%%%%%%%%%%%%%%%%%%%%%%%%%%%%%%%%%%%%%%%%%%%%%%%%%%%%%%%%%%%%%%%%%%%%%
%\section{Introduction}
%\label{sect1}
%%%%%%%%%%%%%%%%%%%%%%%%%%%%%%%%%%%%%%%%%%%%%%%%%%%%%%%%%%%%%%%%%%%%%%%%
%
\section{Introduction}
\noindent
  The past twenty years have shown much progress in the field of
  perturbative calculations in strong interaction physics [1]. This in
  particular holds for the radiative corrections to the deep inelastic
  structure functions. Sometimes these corrections could be even extended
  up to third order in the strong coupling constant $\alpha_s$.
  The structure
  functions we would like to discuss are measured in deep inelastic
  lepton-hadron scattering

%1)                                                                 
\begin{equation}
        l_1 (k_{1}) + H(p) \rightarrow l_{2} (k_{2}) + "X"
\end{equation}

\vspace*{3mm}
\noindent
  where $l_1,l_2$ 
stand for the in- and outgoing leptons respectively. The
  hadron is denoted by $H$ and $"X"$ stands for any inclusive hadronic state.
  The relevant kinematical and scaling variables are defined by
%2)
\begin{equation}
        q = k_1 - k_2 \hspace{1cm}
        q^2 = - Q^2 > 0   \hspace{1cm}
   %1)
     x = \frac{Q^2}{2pq}   \hspace{1cm}
        y = \frac{pq}{pk_1}\\[3mm]
\end{equation}
  with the boundaries
%3)
\begin{equation}
     0 < y < 1  \hspace{3cm}  0 < x \le 1 \\[3mm]
\end{equation}
  Reaction (1) proceeds via the exchange of one of the intermediate vector
  bosons $V$ of the standard model which are represented by 
$V=\gamma,Z,W$.
  In the case of unpolarized scattering with 
$V=\gamma$ one can measure the
  structure functions $F_L(x,Q^2)$ (longitudinal) and 
$F_1(x,Q^2)$ (transverse)
  or the better known $F_2(x,Q^2)$ which is related to the former two via
%4)
\begin{equation}
     F_2 \, (x,Q^2) \, = \, 2x \, F_1 \, (x,Q^2) \, + \,\,F_L
     \,(x,Q^2) \\[3mm]
\end{equation}
  When $V=W$ or $V=Z$ one can in addition to $F_1$, $F_2$
  and $F_L$ also measure 
  the structure function $F_3(x,Q^2)$ which is due to parity violation of
  the weak interactions. In the case the incoming lepton and hadron are 
  polarized one measures besides the structure functions $F_i$ $(i=1,2,3,L)$
  also the spin structure functions denoted by $g_i(x,Q^2)$ $(i=1,\cdots 5)$. 
  At
  this moment, because of the low $Q^2$ available, reaction (1) is only
  dominated by the photon $(V=\gamma)$ so that one has data for $g_1(x,Q^2)$
  (longitudinal spin) and $g_2(x,Q^2)$ (transverse spin) only. 
  The measurement of the structure functions at large $Q^2$ gives us insight
  in the structure of the hadrons. According to the theory of quantum
  chromodynamics (QCD) the hadrons consist out of quarks and gluons where 
  the latter are carriers of the strong force. When $Q^2$ gets large one can
  probe the light cone behaviour of the strong interactions which can be
  described by perturbation theory because the running coupling constant 
  denoted by $\alpha_s(Q^2)$ is small. In
  particular perturbative QCD predicts the $Q^2$-evolution of the deep
  inelastic structure functions mentioned above. Unfortunately the theory
  is not at that stage that it enables us to predict the $x$-dependence so
  that one has to rely on parametrizations which are fitted to the data.
  A more detailed description of the structure functions is provided by
  the parton model which can be applied if one can neglect power corrections
  of the type $(1/Q^2)^p$ (higher twist effects). Here one asumes that in the
  Bjorken limit $(Q^2 \rightarrow \infty$, $x$ is fixed) 
 the interaction between the hadron
  and lepton in process (1) proceeds via the partons (here the quarks and
  the gluons) of the hadron. If the scattering of the lepton with the partons
  becomes incoherent the structure function can be written as

%5)
\begin{eqnarray}
F^{V,V'} (x,Q^2) & = & \int^1_x \,\frac{dz}{z} \, \left[
\sum^{n_f}_{k=1}\,
\left(v ^{(V)}_k \, v^{(V')}_k \,+\, a^{(V)}_k \,
a^{(V')}_k \right) \,
\left\{ \Sigma  (\frac{x}{z}, \mu^2) \, C^S_{i,q} \,
(z, \frac{Q^2}{\mu^2})  \right. \right.
\nonumber \\
& + &   \left. \left.
G\, \left(\frac{x}{z}, \mu^2 \right) \, C_{i,g} \,
\left( z, \frac{Q^2}{\mu^2} \right) \right\} \,
+ \sum^{n_f}_{k=1} \, \left( v^{(V)}_k \,
v^{(V')}_k \, + \,a^{(V)}_k
\, a^{(V')}_k \right) \right. \nonumber \\
& \Delta_k & \left. \left(\frac{x}{z}, \mu^2 \right) \,
C^{NS}_{i,q}  \, \left( z, \,\frac{Q^2}{\mu^2} \right) \right]
\hspace{2cm} i = 1, 2, L
\end{eqnarray}

%6)
\begin{eqnarray}
F^{V,V'}_3 \left( x,Q^2 \right) & = & \int^1 _x \,
\frac{dz}{z} \, \left[ \sum^{n_f}_{k=1} \,
\left( v^{(V)}_k a^{(V')}_k \, + \, a^{(V)}_k \,
v^{(V')}_k \right) \right. \nonumber \\
& V_k & \left. \left( \frac{x}{z}, \mu^2 \right) \,
C ^{NS}_{3,q} \, \left( z,
\frac{Q^2}{\mu^2} \right) \right]
\end{eqnarray}

\noindent
  with similar expressions for the twist two contributions to the spin 
  structure functions $g_1(x,Q^2)$ in which case we introduce the notations
  $\Delta \Sigma, \Delta G , \Delta C_{i,l}$ etc.. The vector- and axial-vector
  electroweak couplings of the standard model are given by $v_k^{(V)}$ and 
  $a_k^{(V)}$ respectively with $V=\gamma,Z,W$ and $k=1 (u), 2 (d), 3
  (s)$
 .... .
  Further $n_f$ denotes the number of light flavours and $\mu$ stands for the
  factorization/renormalization scale. The singlet $(\Sigma)$ and non-singlet
  combinations of parton densities $(\Delta_k, V_k)$ are defined by

%7)
\begin{equation}
\Sigma \left( z, \mu^2 \right) = \frac{1}{n_f} \sum_{k=1}^{n_f}
\left( f_k \left( z, \mu^2 \right)
+ f_{\bar{k}} \left( z, \mu^2 \right) \right)
\end{equation}

%8)
\begin{equation}
\Delta_k \, (z, \mu^2)\,=\, f_k\,(z, \mu^2)\, +\, f_{\bar{k}}\, (z,
\mu^2) \,-\, \Sigma \,(z, \,\mu^2)
\end{equation}

%9)
\begin{equation}
V_k\,(z, \,\mu^2)\, =\, f_k \,(z, \,\mu^2) - f_{\bar{k}} \,(z,\,
\mu^2)
\end{equation}
  
  where $f_k , f_{\bar k}$ denote the quark and anti-quark densities of species
  $k$ respectively. The gluon density is defined by $G(z,\mu^2)$. The same
  nomenclature holds for the coefficient functions $C_{i,l} (l=q,g)$ which can
  also be distinguished in a singlet (S) and a non-singlet (NS) part. Like
  in the case of the structure functions the $x$-dependence of the parton
  densities cannot be determined by perturbative QCD and it has to be
  obtained by fitting the parton densities to the data. Fortunately these
  densities are process independent and they are therefore universal. This
  property is not changed after including QCD radiative corrections. It
  means that the same parton densities also show up in other so called hard
  processes like jet production in hadron-hadron collisions,direct photon 
  production, heavy flavour production, Drell-Yan process etc. Another
  firm prediction of QCD is that the scale $(\mu)$ evolution of the parton
  densities is determined by the DGLAP [2] splitting functions $P_{ij}$
  $(i,j=q,\bar q,g)$ which can be calculated order by order in the strong 
  coupling constant $\alpha_s$. The perturbation series of $P_{ij}$
 gets the form

%10)
\begin{equation}
P_{kl} = a_s \, P^{(0)}_{kl} + a^2_s \,P^{(1)}_{kl} + a^3_s \,
P^{(2)}_{kl} + .\
.. \\[3mm]
\end{equation}
  with $a_s=\alpha_s(\mu^2)/{4\pi}$. The splitting functions $P_{ij}$
  are related
  to the anomalous dimensions $\gamma_{ij}^{(n)}$ 
corresponding to twist two local
  operators $O_i^{\mu_1 ...\mu_n} (x)$ of spin $n$ via the Mellin transform
%11)
\begin{equation}
\gamma^{(n)}_{ij} = - \int^1 _o dz  z^{n-1} \, P_{ij} (z) \\[3mm]
\end{equation}
  These operators appear in the light cone expansion of the product of two 
  electroweak currents which shows up in the calculation of the cross section
  of process (1)
%12)
\begin{equation}
J(x)\, J(0) \sim   \sum^{\infty}_{n=0} \sum_k \,\widetilde{C}^{(n)}_k
\,
(\mu^2x^2) \, x_{\mu{_1}} .. x_{\mu{_n}} \, O_k^{\mu_1 ...\mu_n} (0,
\mu^2)  \\[3mm]
\end{equation}
  where $\widetilde {C}_k^{(n)}$ (12) are the Fourier transforms of
  the coefficient
  functions $C_k^{(n)}$ (5),(6) $(k=q,g)$ in Minkowski space
  $(x_\mu)$. 
  Like the
  splitting functions they are calculable order by order in $\alpha_s$ and the 
  perturbation series takes the form
%13)
\begin{equation}
C_{i,k} = \delta_{kq} + a_s \, C^{(1)}_{i,k} + a_s^2 \,C^{(2)}_{i,k} +
 a_s^3 \, C^{(3)}_{i,k} + ...  \\[3mm]
\end{equation}
  with $i=1,2,3,L$ and $k=q,g$.
  We will now review the higher order QCD corrections to the splitting 
  functions and the coefficient functions which have been calculated  
  till now.

\section {Splitting Functions}
The splitting functions are calculated by
\begin{itemize}
\item[1.] $P_{ij}^{(0)}$\hspace*{1cm}      Gross and Wilczek (1974) [3];
                         Altarelli and Parisi (1977) [2].
\item[2.] $\Delta P_{ij}^{(0)}$\hspace*{1cm}       Sasaki (1975) [4];
                         Ahmed and Ross (1976) [5];
                         Altarelli and Parisi  [2].
\item[3.] $P_{ij}^{(1)}$\hspace*{1cm} Floratos, Ross, Sachrajda (1977) [6];
                         Gonzales-Arroyo, Lopez, Yndurain (1979)
                 \hspace*{15mm} [7];
                         Floratos, kounnas, Lacaze (1981) [8];
                         Curci, Furmanski, Petronzio (1980) [9].
\item[4.] $\Delta P_{ij}^{(1)}$\hspace*{1cm}  Zijlstra and van Neerven (1993)
          [10];
                         Mertig and van Neerven (1995) [11];
                   \hspace*{2cm}      Vogelsang (1995) [12].
\end{itemize}

  Notice that till 1992 there was a discrepancy for $P_{gg}^{(1)}$ between
  the covariant gauge [6--8] and the lightlike axial gauge calculation
  [9] which was decided in favour of the latter by Hamberg and van Neerven
  who repeated the covariant gauge calculation in [12]. The DGLAP splitting
  functions satisfy some special relations. The most interesting one is
  the so called supersymmetric relation which holds in ${\cal N} =1$ 
  supersymmetry [13]. Here the colour factors, which in $SU(N)$ are given by
  $C_F=(N^2-1)/{2N}$ , $C_A=N$, $T_f=1/2$  become $C_F=C_A=2T_f=N$. The
  supersymmetric relation then reads
%14)
\begin{equation}
P^{S,(k)}_{qq} + P^{(k)}_{gq} - P^{(k)}_{qg} - P^{(k)}_{gg} = 0 \\[3mm]
\end{equation}
%15)
\begin{equation}
\Delta P^{S,(k)}_{qq} + \Delta P^{(k)}_{gq} - \Delta P^{(k)}_{qg}
- \Delta P^{(k)}_{gg} = 0  \\[3mm]
\end{equation}
  which is now confirmed up to first $(k=0)$ and second $(k=1)$ order in 
  perturbation theory.
  The third order splitting functions $P_{ij}^{(2)},\Delta
  P_{ij}^{(2)}$
  are not known
  yet. However the first few moments $\gamma_{ij}^{(2),(n)}$ for $n=2,4,6,8,10$
  have been calculated by Larin, van Ritbergen, Vermaseren (1994) [14].
  Besides exact calculations one has also determined the splitting functions
  and the anomalous dimensions in some special limits. Examples are the
  large $n_f$ expansion carried out by Gracey (1994) [15]. 
  Here one has computed the 
  coefficients $b_{21}$ and $b_{31}$ in the perturbations series of the 
  non-singlet
  anomalous dimension
%16)
\begin{eqnarray}
\left. \gamma^{NS}_{qq} \right|_{n_f \rightarrow \infty }
& = & a_s^2 \, \left[ n_f \, C_F \, b_{21} \right]
+ a_s^3\, \left[ n^2_f \, C_F \, b_{31}  \right. \nonumber \\
& + & \left. n_f \, C_A \, C_F \, b_{32}
 +   n_f \, C_F^2 \, b_{33} \right] + ...
\end{eqnarray}\\[2ex]
  Further Catani and Hautmann (1993) [16] calculated the splitting functions
  $P_{ij}(x)$ in the limit $x \rightarrow 0$. The latter take the following 
  form
\vspace*{3mm} \noindent
%17
\begin{equation}
\left. P^{(k)}_{ij} (x) \right|_{x \rightarrow 0}
 \sim \, \frac{ln^kx}{x} \rightarrow
\left. \gamma^{(k),(n)}_{ij} \right|_{n \rightarrow 1}
 \sim \, \frac{1}{(n-1)^{k+1}} \\[3mm]
\end{equation}
  The above expressions follow from the BFKL equation [17] and 
  $k_T$-factorization [18]. Some results are listed below. The leading
  terms in $\gamma_{gg}^{(n)}$ are given by
%18)
\begin{equation}
\left. \gamma^{(n)}_{gg}\right|_{n \rightarrow 1} = \left[ C_A \frac{a_s}{n-1} 
\right] + 2\zeta(3) \,
\left[ C_A
\frac{a_s}{n-1} \right] ^4 + 2\zeta(5) \, \left[ C_A \frac{a_s}{n-1}
\right] ^6   \\[3mm]
\end{equation}
  where $\zeta(n)$ denotes the Riemann zeta-function. Further we have in
  leading order $1/(n-1)$

%19)
\begin{equation}
\left. \gamma^{(n)}_{gq}\right|_{n \rightarrow 1} = \frac{C_F}{C_A} \,\left.
 \gamma^{(n)}_{gg}\right|_{n \rightarrow 1} 
\end{equation}
%\vspace*{3mm} \noindent
%20)
\begin{eqnarray}
\left. \gamma^{(n)}_{qg} \right|_{n \rightarrow 1} & = & a_s T_f \frac{1}{3}\, 
\left[ 1 + 1.67 \,
\left\{ \frac{a_s}{n-1} \right\} \, + \, 1.56 \,
\left\{ \frac{a_s}{n-1} \right\} ^2 \right. \nonumber \\
& + & 3.42 \left. \left\{ \frac{a_s}{n-1} \right\} ^3 \, + \, 5.51
\left\{ \frac{a_s}{n-1} \right\} ^4 \, + ...\right] 
\end{eqnarray}
%21)
\begin{equation}
\left. \gamma^{S,(n)}_{qq}\right|_{n \rightarrow 1} = \frac{C_F}{C_A} \, 
\left[\left. \gamma^{(n)}_{qg} \right|_{n \rightarrow 1} -
\frac{1}{3} \,  a_s T_f \right]  \\[3mm]
\end{equation}\\[3mm]
  Kirschner and Lipatov (1983) and Bl\"umlein and Vogt (1996) 
  have also determined the subleading terms
  in the splitting functions (anomalous dimensions). They behave like
%22)
\begin{equation}
\left. P^{(k)}_{ij} (z) \right|_{z \rightarrow 0}
 \sim \, ln^{2k}\,z
\left. \hspace{2cm} \gamma^{(k),(n)}_{ij}
\right|_{n \rightarrow 0}
\sim \, \frac{1}{n^{2k+1}}
\end{equation}\\[3mm]
  The same logarithmic behaviour also shows up in $\Delta P_{ij}$ and $\Delta$
  $\gamma_{ij}^{(n)}$. In the latter case the expressions in (22) become the
  leading ones since the most singular terms in (17) decouple in the spin
  quantities. The expressions in (22) have been calculated for the spin
  case by Bartels, Ermolaev, Ryskin (1995) [20] and by Bl\"umlein and
  Vogt (1996) [21] who also investigated the effect of these type of
  corrections on the spin structure function $g_1(x,Q^2)$.
  Finally the three-loop anomalous dimension $\Delta \gamma_{qq}^{S,(1)}$
  is also known (see Chetyrkin,K\"uhn (1993) [22] and Larin (1993) [23]).
  It reads
%23)
\begin{equation}
\Delta \gamma^{S,(1)}_{qq} = a^2_s \,
\left[ -6n_f \, C_F \right] + a^3_s \,
\left[ \left( 18 C^2_F -
\frac{142}{3} C_AC_F \right) n_f + \frac{4}{3}\, n^2_f C_F \right] \\[3mm]
\end{equation}
  Notice that the second order coefficient was already determined by Kodaira
 (1980) [24].

\section {Coefficient Functions}
\noindent
The higher order corrections to the coefficient functions are calculated by
\begin{itemize}
\item[1.] $C_{i,q}^{(1)}$ , $C_{i,g}^{(1)}$~~~$i=1,2,3,L$ \hspace*{1cm}  
Bardeen, Buras, 
Muta, Duke (1978) [25],\\
\hspace*{6cm} see also Altarelli (1980) [26].

\item[2.]  $\Delta C_q^{(1)}$ , $\Delta C_g^{(1)}$ \hspace*{2cm}   
Kodaira et al. (1979) [27],\\
\hspace*{5cm}    see also Anselmino, Efremov, Leader 
                                     (1995) [28]
\end{itemize}
  Together with the splitting functions $P_{ij}^{(k)}$, $\Delta 
  P_{ij}^{(k)}\,(k=0,1)$
  one is now able to make a complete next-to-leading (NLO) analysis of the
  structure functions $F_i(x,Q^2)\, (i=1,2,3,L)$ and $g_1(x,Q^2)$. The second
  order contributions to the coefficient functions are also known
\begin{itemize}
\item[1.]  $C_{i,q}^{(2)} , C_{i,g}^{(2)}$~~~ $i=1,2,3,L$ \hspace*{1cm} 
Zijlstra and van Neerven (1991) [29]
\item[2.]  $\Delta C_q^{(2)} , \Delta C_g^{(2)}$ \hspace*{3cm}   Zijlstra and 
van Neerven (1993) [10]
\end{itemize}

  The first few moments of $C_{i,k}^{(2)}$~~$(i=2,L ; k=q,g)$ were calculated 
by
  Larin and Vermaseren (1991) [30] and they agree with Zijlstra and van
  Neerven [29]. The first moment of $\Delta C_q^{(2)}$ was checked by Larin 
  (1993) [31] and it agrees with the result of Zijlstra and van Neerven
  [10]. The third order contributions to the coefficient functions are
  not known except for some few moments. They are given by
\begin{itemize}
\item[1.]  $C_{1,q}^{(3),(1)}$  (Bjorken sum rule) \hspace*{3cm}  Larin, 
Tkachov, 
  Vermaseren (1991) [32];
\item[2.]  $C_{3,q}^{(3),(1)}$  (Gross-Llewellyn
                    Smith sum rule)\hspace*{1cm} Larin and Vermaseren (1991) 
[33];
\item[3.]  $\Delta C_q^{(3),(1)}$ (Bjorken sum
                         rule )\hspace*{3cm}  Larin and Vermaseren (1991) [33];
\item[4.]  $C_{i,q}^{(3),(n)}$~~ $(i=2,L)~~ n=2,4,6,8$ \hspace*{15mm}
    Larin, van Ritbergen, Vermaseren (1994) [14],\\
\hspace*{8cm}                                          (see also [34]).
\end{itemize}

  Since the three-loop splitting functions $P_{ij}^{(2)}$, $\Delta 
  P_{ij}^{(2)}$ are
  not known, except for a few moments, it is not possible to obtain a
  full next-to-next-to-leading order (NNLO) expression for the structure
  functions. However recently Kataev et al. (1996) [35] made a NNLO analysis
  of the structure functions $F_2(x,Q^2)$, $F_3(x,Q^2)$ (neutrino scattering)
  in the kinematical region $x > 0.1$ which is based on 
  $\gamma_{qq}^{NS,(2),(n)}$
  for $n=2,4,6,8,10$ [14].
  Like in the case of the DGLAP splitting functions Catani and Hautmann (1994)
  [16]
  also derived the small $x$-behaviour of the coefficient functions.
  At small $x$ the latter behave like

%24)
\begin{equation}
\left. C^{(l)}_{i,k} \right|_{x \rightarrow 0}
\sim \, \frac{ln^{l-2}x}{x} \hspace{2cm}
\left. C^{(l),(n)}_{i,k} \right|_{n \rightarrow 1}
\sim \, \frac{1}{(n-1)^{l-1}} \hspace*{1cm} (l \ge 2)
\end{equation}[3mm]
  The ingredients of the derivation are again the BFKL equation [17] and
  $k_T$-factorizaton [18]. from [16] we infer the following Mellin-transformed
  coefficient functions.

%25)
\begin{eqnarray}
\left. C^{(n)}_{L,g}\right|_{n \rightarrow 1} & = & a_s \, T_f \, n_f \,
\frac{2}{3} \, \left[ 1 - 0.33 \,
\left\{ \frac{a_s}{n-1} \right\} \,
+ \, 2.13 \, \left\{ \frac{a_s}{n-1} \right\}^2 \right. 
\nonumber \\
& + & \left.
2.27 \left\{ \frac{a_s}{n-1} \right\} ^3 \, + \,
0.43 \, \left\{ \frac{a_s}{n-1} \right\} ^4 \, + \, ... \right] 
\end{eqnarray}
%26)
\begin{eqnarray}
\left. C^{(n)}_{2,g}\right|_{n \rightarrow 1} & = & a_s \, T_f \, n_f \, 
\frac{1}{3}
\, \left[ 1 + 1.49 \,
\left\{ \frac{a_s}{n-1} \right\}
\, + \, 9.71 \, \left\{ \frac{a_s}{n-1} \right\}^2 \right. \nonumber
\\
& + & \left. 16.43 \, \left\{ \frac{a_s}{n-1} \right\}^3 \, +
\, 39.11 \, \left\{ \frac{a_s}{n-1} \right\} ^4 \, + ... \right]
\end{eqnarray}
%27)
\begin{equation}
\left. C^{S,(n)}_{L,q}\right|_{n \rightarrow 1} = \frac{C_F}{C_A} \, \left[ 
\left. C^{(n)}_{L,g}\right|_{n \rightarrow 1} -
\frac{2}{3} \, a_s
\, n_f \, T_f \right]  \\[3mm]
\end{equation}
%28)
\begin{equation}
\left. C^{S,(n)}_{2,q}\right|_{n \rightarrow 1} = \frac{C_F}{C_A} \, \left[ 
\left. C^{(n)}_{2,g} \, \right|_{n \rightarrow 1} - \,
\frac{1}{3} \, a_s \, n_f  \, T_f \right] \\[3mm]
\end{equation}             
  The order $\alpha_s^2$ coefficients were already obtained via the exact 
  calculation performed by Zijlstra and van Neerven (1991) [29]. The 
  subleading terms given by
%29)
\begin{equation}
\left. C^{(l)}_{2,k} \right|_{x \rightarrow 0}
\sim \ln ^{2l-1} x \hspace{2cm} (l \ge 1) \\[3mm]
\end{equation}
  were investigated by Bl\"umlein and Vogt (1996) [21]. The most singular
  terms shown in (24) do not appear in the spin coefficient functions
  $\Delta C_k^{(l)}$ because the Lipatov pomeron decouples in polarized 
  lepton-hadron scattering. Therefore the most singular behaviour near $x=0$ 
  is given by (29) (see [10],[21]). Besides the logarithmical enhanced
  terms which are characteristic of the low $x$-regime we also find similar
  type of logarithms near $x=1$. Their origin however is completely different 
  from the one determining the small $x$-behaviour. The logarithmical
  enhanced terms near $x=1$, which are actual distributions, originate from
  soft gluon radiation. They dominate the structure functions $F_i$ and $g_i$ 
  near $x=1$ because other production mechanisms are completely suppressed
  due to limited phase space. Following the work in [36] and [37] the
  DGLAP splitting functions and the coefficient functions behave near $x=1$
  like
%30)
\begin{equation}
P^{NS,(k)}_{qq} = \Delta \, P^{NS,(k)}_{qq} \sim \left( \frac{1}{1-x}\right)_+
\hspace{1cm}
 P^{(n)}_{gg}  =
\Delta P^{(k)}_{gg} \sim \, \left( \frac{1}{1-x}\right)_+
\end{equation}
%31)
\begin{equation}
 \Delta C^{NS,(k)}_q  =
C^{NS,(k)}_{i,q} \sim \left( \frac{\ln^{2k-1} (1-x)}{1 - x} \right)_+
  \hspace{1cm} (l = 1, 2, 3) \\[3mm]
\end{equation}
  Notice that the above corrections cannot be observed in the kinematical 
  region $(x < 0.4)$ accessible at HERA. Furthermore the behaviour in (30)
  is a conjecture (see [7]) which is confirmed by the existing calculations 
  carried out up to order $\alpha_s^2$.

\section {Heavy Quark Coefficient Functions}
\noindent
The heavy quark coefficient functions have been calculated by
\begin{itemize}
\item[1.] $C_{i,g}^{(1)}(x,Q^2,m^2)$~~$(i=2,L)$\hspace*{1cm} Witten (1976) 
[38];
\item[2.] $\Delta C_g^{(1)}(x,Q^2,m^2)$ \hspace*{2cm}     Vogelsang (1991) 
[39];
\item[3.] $C_{i,g}^{(2)}(x,Q^2,m^2),$~~ $C_{i,q}^{(2)}(x,Q^2,m^2)$~~$(i=2,L)$  
\hspace*{5mm}   Laenen, Riemersma, Smith, van Neerven
\hspace*{9cm} (1992) [40].
\end{itemize}
  where $m$ denotes the mass of the heavy quark. The second order heavy quark
  spin coefficient functions $\Delta C_g^{(2)}(x,Q^2,m^2)$ and 
  $\Delta C_q^{(2)}(x,Q^2,m^2)$ are not known yet. Due to the presence of the
  heavy quark mass one was not able to give explicit analytical expressions 
  for $C_{i,k} (i=2,L ; k=q,g)$. However for experimental and 
  phenomenological use 
  they were presented in the form of tables in a computer program [41].
  Analytical expressions do exist when either $x \rightarrow 0$ or 
  $Q^2 \gg m^2$. In the
former case Catani, Ciafaloni and Hautmann [42] derived the general form

%32)
\begin{equation}
\left. C^{(l)}_{i,k} \right|_{x \rightarrow 0}
\sim   \,  \frac{1}{x} \, \ln^{l-2} (x) \, f(Q^2, m^2)
\hspace{1cm} (l \ge 2,\, i = 2,\, L; \,  k=q,g) \\[3mm]
\end{equation}
  Like for the light parton coefficient functions (see (24)) the above
  expression is based on the BFKL equation [17] and $k_T$-factorization [18].
  In second order Buza et al. (1996) [43] were able to present analytical
  formulae for the heavy quark coefficient functions in the asymptotic
  limit $Q^2 \gg m^2$. This derivation is based on the operator product
  expansion and mass factorization.

  \section {Phenomenology at low $x$}
  \noindent

  Since the calculation of the higher order corrections to the DGLAP
  splitting functions $P_{ij}$ and the coefficient functions $C_{ik}$ is very
  cumbersome various groups have tried to make an estimate of the NNLO 
  corrections to structure functions in particular to $F_2(x,Q^2)$. The
  most of these estimates concerns the small $x$-behaviour. In [44]
  Ellis, Kunszt and Levin Hautman made a detailed study 
  of the $Q^2$-evolution
  of $F_2$ using the small $x$-approximation for $P_{ij}$ (17) and
  $C_{2,k}^{(2)}$ (24).
  Their results heavily depend on the set of parton densities used and
  the non leading small $x$-contributions to $P_{ij}^{(2)}$. The latter are 
e.g. 
  needed to satisfy the momentum conservation sum rule condition. Large
  corrections appear when for $x \rightarrow 0$ the gluon density behaves like
  $xG(x,\mu^2)\rightarrow$
  const. whereas
  they are small when the latter has the behaviour $xG(x,\mu^2) \rightarrow
  x^{-\lambda} (\lambda
  \sim 0.3 - 0.5$ ; Lipatov pomeron).\\
  However other investigations reveal that
  the singular terms at $x=0$, present in $P_{ij}$ and $C_{i,k}$, 
  do not dominate
  the radiatve corrections to $F_2(x,Q^2)$ near low $x$. This became apparent
  after the exact coefficient functions or DGLAP splitting functions were
  calculated.\\
  In [45] Gl\"uck, Reya and Stratmann (1994) investigated the 
  singular behaviour of the second order heavy quark coefficient functions
  (32) in
  electroproduction and they found that its effect on $F_2$ was small.\\
  Similar work was done by Bl\"umlein and Vogt (1996) [21] on the effect of
  the logarithmical terms (22),(29) on $g_1(x,Q^2)$ which contribution to
  the latter turned out to be negligable.\\
  Finally we would like to illustrate
  the effect of the small $x$-terms, appearing in the coefficient functions 
  $C_{2,k}^{(2)}$
  and $C_{L,k}^{(2)}$, on the structure functions $F_2(x,Q^2)$ and 
$F_L(x,Q^2)$.
  For that purpose we compute the order $\alpha_s^2$ contributions to $F_2$ 
  and $F_L$. Let us introduce the following notations. When the exact
  expressions for the coefficient functions $C_{i,k}^{(2)}$ are adopted 
  the order
  $\alpha_s^2$ contributions to $F_i$ will be called $\delta F_i^{(2),exact}$.
  If we replace the exact coefficient functions by their most singular
  part which is proportional to $1/x$ (see (24)) the order $\alpha_s^2$ 
  contributions to $F_i$ are denoted by $\delta F_i^{(2),app}$. The results
  are listed in table 1 and 2 below. Further we have used the parton density
  sets MRS(D0) $(xG(x,\mu^2) \rightarrow$ $const.$  for $x \rightarrow 0)$
  and MRS(D-) $(xG(x,\mu^2) \rightarrow x^{-\lambda}$ for $x
\rightarrow 0)$ [46]
\vspace*{2cm}

\begin{tabular}{||l||r|r|r||r|r|r||}
\hline \hline
\multicolumn{1}{||c|}{  }&
\multicolumn{3}{ c|}{MRS(D0)}&
\multicolumn{3}{c||}{MRS(D-)}\\
\multicolumn{1}{||c|}{$x$}&
\multicolumn{1}{||c|}{$F_2^{\rm NLO}$}&
\multicolumn{1}{c|}{$ \delta F_2^{(2),exact}$}&
\multicolumn{1}{c|}{$ \delta F_2^{(2),app}$}&
\multicolumn{1}{|c|}{$F_2^{\rm NLO}$}&
\multicolumn{1}{c|}{$ \delta F_2^{(2),exact}$}&
\multicolumn{1}{c||}{$ \delta F_2^{(2),app}$}\\
\hline \hline
$10^{-3}$ & 0.67 & -0.069 & 0.088 & 0.99 & -0.084 & 0.116 \\
$10^{-4}$ & 0.82 & -0.088 & 0.158 & 2.29 & -0.226 & 0.349 \\
$10^{-5}$ & 1.00 & -0.092 & 0.251 & 5.99 & -0.665 & 1.059 \\
\hline\hline
\multicolumn{1}{||c|}{$x$}&
\multicolumn{1}{||c|}{$F_L^{\rm NLO}$}&
\multicolumn{1}{c|}{$ \delta F_L^{(2),exact}$}&
\multicolumn{1}{c|}{$ \delta F_L^{(2),app}$}&
\multicolumn{1}{|c|}{$F_L^{\rm NLO}$}&
\multicolumn{1}{c|}{$ \delta F_L^{(2),exact}$}&
\multicolumn{1}{c||}{$ \delta F_L^{(2),app}$}\\
\hline\hline
$10^{-3}$ & 0.149 & -0.029 & -0.040 & 0.263 & 0.008 & -0.052 \\
$10^{-4}$ & 0.210 & -0.062 & -0.071 & 0.780 & 0.031 & -0.156 \\
$10^{-5}$ & 0.281 & -0.102 & -0.113 & 2.370 & 0.105 & -0.475 \\
\hline\hline
\end{tabular}

\vspace*{2cm} \noindent

  {}From the table above we infer that a steeply rising gluon density near 
  $x=0$ (MRS(D-)) leads to small corrections to $F_2$ and $F_L$. 
On the other hand
  if one has a flat gluon density (MRS(D0)) the corrections are much larger
  in particular for $F_L$. A similar observation was made for $F_2$ in [44].
  However the most important observation is that the most singular part
  of the coefficient functions gives the wrong prediction  for the order
  $\alpha_s^2$ contributions to the structure functions except for $F_L$ 
  provided the set MRS(D0) is chosen. This means that the subleading terms
  are important and they cannot be neglected. Therefore our main conclusion 
  is that only exact calculations provide us with the correct NNLO analysis 
  of the structure functions. The asymptotic expressions obtained in the 
  limits $x \rightarrow 0 , x \rightarrow 1$ and $Q^2 \gg m^2$ can 
  only serve as a check on the
  exact calculations of the DGLAP splitting functions and the coefficient
  functions.

%%%%%%%%%%%%%%%%%%%%%%%%%%%%%%%%%%%%%%%%%%%%%%%%%%%%%%%%%%%%%%%%%%%%%%%%

\end{document}